\documentstyle[12pt,aaspp4]{article}        % Large preprint style
\received{{23 August 1996}}
\cpright{AAS}{1997}
\lefthead{Hjorth \& Tanvir}
\righthead{Fundamental plane in \leo\ and H_0}

\slugcomment{To appear in ApJ 1997}

\newcommand{\etal}{et al.}

\newcommand{\paper}{paper}
\newcommand{\kms}{\ {\rm km\,s}^{-1}}
\newcommand{\mg}{{\rm Mg_2}}
\newcommand{\leo}{{\rm Leo-I}}
\renewcommand{\rq}{$R^{1/4}$}
\newcommand{\ho}{\kms\ {\rm Mpc}^{-1}}
\begin{document}

\title{{Calibration of the fundamental plane zero-point in the Leo-I group and 
        an estimate of the Hubble constant}\footnote{Based on observations 
        made with INT operated on the island of La Palma by the Royal 
        Greenwich Observatory in the Spanish Observatorio del Roque de los 
        Muchachos of the Instituto de Astrof\'\i sica de Canarias.}}

\author{Jens Hjorth\altaffilmark{2} and Nial R. Tanvir}

\altaffiltext{2}{Present address: NORDITA, Blegdamsvej 17, DK--2100 Copenhagen
\O, Denmark; jens@nordita.dk.}

\affil{Institute of Astronomy, Madingley Road, Cambridge CB3 0HA, UK}

\authoraddr{Jens Hjorth,Institute of Astronomy, Madingley Road, 
            Cambridge CB3 0HA, UK}

\begin{abstract}

We derive new effective radii and total magnitudes for 5 E and S0 galaxies in 
the \leo\ group from wide-field CCD images. These are used in conjunction with 
recent literature velocity data to construct the fundamental plane (FP) of 
the \leo\ group. The rms scatter that we find is only 6 \% in distance. The 
zero point of this relation provides a calibration of the FP as a distance 
indicator and directly determines the angular diameter distance ratio between 
the \leo\ group and more distant clusters. In the language of Jerjen and 
Tammann (1993) we determine a cosmic velocity of the \leo\ group of 
$757\pm 68 \kms$ relative to the Coma cluster, or $796\pm 57 \kms$ 
relative to a frame of 9 clusters. Combining this velocity with the Cepheid 
distance to M96, a member of \leo, we find the Hubble constant to be 
$H_0=67\pm 8\ho$ or $H_0=70\pm7\ho$ for each case.  The distance we obtain 
for the Coma cluster itself ($108\pm12$ Mpc) is in good agreement with a 
number of other recent estimates.

\end{abstract}


\section{INTRODUCTION}

Despite the successful determination of Cepheid distances to several spiral 
galaxies in the Virgo cluster and elsewhere with the Hubble Space Telescope 
(HST), the dispute over the value of the Hubble constant is still not settled 
(Tanvir \etal\ 1995; Mould \etal\ 1995; Sandage \etal\ 1996). Part of the 
problem undoubtedly lies with the uncertain peculiar velocity and significant 
depth, particularly for the late-type galaxies, of the Virgo cluster. These 
uncertainties are removed if secondary indicators can be used to extend 
distance measurements directly to more remote and, ideally, more compact 
clusters. Moreover, due to the morphological segregation of early type 
galaxies towards the cores of clusters, secondary distance indicators based 
on E and S0 galaxies are preferred.

In this \paper\ we calibrate the so-called fundamental plane (FP) distance 
indicator (Djorgovski \& Davis 1987; Dressler \etal\ 1987) in the \leo\ 
group. The \leo\ group (Ferguson \& Sandage 1990) is the nearest group of 
galaxies containing both bright spirals as well as early type galaxies. 
The center of the group is defined by the E1 galaxy NGC 3379 \mbox{(= M105)} 
and the S0 galaxy NGC 3384 which form a close pair. Planetary nebula 
luminosity function (PNLF) and surface brightness fluctuations (SBF) 
observations (Ciardullo, Jacoby, \& Tonry 1993) indicate that the E5 galaxy 
NGC 3377 is a member of the group, whilst the radial velocity and projected 
distance of the S0 galaxy NGC 3412 from the core of the group indicate that 
it too is a likely member (Garcia 1993; Tanvir 1996). Finally, the S0/a 
galaxy NGC 3489 is also a possible early-type member of \leo, however, given 
its distance from the group center ($\sim3^{\circ}$) and its comparatively 
``late-type'' properties, we decided {\it a priori} that it should not be 
included in the construction of the fundamental plane (see \S 3).

The fundamental plane is an excellent distance indicator relating distances 
between E and S0 galaxies particularly in clusters and groups. The most recent
paper on the subject studies the FP for 230 E and S0 galaxies in 11 clusters 
and groups (J\o rgensen, Franx \& Kj\ae rgaard 1996; hereafter JFK). For the 
present work, the important result of that study is the demonstration that a 
single global FP yields unbiased distance estimates to early-type galaxies in 
very different environments with a formal intrinsic scatter of 14 \% in 
distance per galaxy. Thus with four early-type galaxies in the \leo\ group we 
may obtain the zero-point of the relationship to 7\%, with the additional 
uncertainty of the absolute distance to the group.

The distance to \leo\ has recently been measured by Cepheids in the spiral 
galaxies M96 (Tanvir \etal\ 1995) and M95 (Graham \etal\ 1997), by SBF 
(Tonry \etal\ 1997; Sodemann and Thomsen 1995; 1996), by the tip of the 
red-giant branch (TRGB) in NGC 3379 (Sakai \etal\ 1997) and PNLF in several 
group members (Ciardullo, Jacoby and Ford 1989; Feldmeir, Ciardullo, \&
Jacoby 1997). Thus 
the \leo\ group is an 
important stepping stone in the determination of the Hubble constant.

\section{GALAXY PARAMETERS}

\subsection{Observations and image analysis}

Observations of the 5 E and S0 galaxies in the \leo\ group were carried out in
the $VRI$ passbands using the INT prime focus camera on 1994 March 14. The CCD 
employed was the thick $1280\times 1180$ EEV5 chip with 0\farcs 557 pixels. 
Conditions were photometric with reasonable seeing of FWHM 1\farcs3. High 
S/N twilight flats were obtained in the morning and evening and standard stars 
were observed through the night to determine color terms and photometric zero 
points. Magnitudes were measured in $\approx$ 12 arcsec diameter apertures, 
similar to those used by Landolt (1992).

In this \paper\ we discuss the $R$ band data obtained. 
The transformation onto the Cousins/Landolt system was
derived from 28 observations of 10 Landolt (1992) fields each of 
which contains several standard stars. A total of 86 independent flux 
measurements went into the derivation of the color term
and these are  plotted in Figure~1. As can be seen, the 
detector and filter used were very well matched to the standard 
system, as no measurable color term or non-linearity was found 
for stars over a large color range $-0.2<V-{R_C}<1.1$
(a typical E galaxy
has $V-{R_C}\approx0.5$). The rms scatter around the linear fit is 0.02 mag, 
most of which can be 
accounted for by known observational errors (photon noise and 
CCD read noise) combined with the quoted uncertainty of the Landolt (1992) 
magnitudes.

The photometric zero-point was determined from the
observations of a subset of 11 standard 
stars which were interspersed with the galaxy observations. Aperture 
corrections were determined from the frames out to $\approx$ 33 arcsec 
diameter apertures and were found to be small. All observations, of both 
standard stars and target galaxies, were obtained at less than 1.15 air masses.
Thus the extinction corrections, for which we used the Carlsberg Meridian 
Circle value for that night of $A_R=0.07$ (converted from $A_V=0.14$ using 
the standard La Palma extinction curve), were very small.

On-target exposures ranged from 10 to 40 secs, chosen
to ensure that the images did not saturate in the galaxy
cores.  Two or three exposures were obtained in each filter
for all the galaxies.
Surface photometry was performed using the profile fitting software developed 
by B. Thomsen (Sodemann \& Thomsen 1994, 1995; Hjorth \etal\ 1995). A mean
growth curve in circular annuli was derived from a 6th order harmonic 
expansion fit to each galaxy in 40 rings, equidistant in \rq. For the small 
galaxies the sky value was determined from the corners of the frames and so 
gives precise total magnitudes and effective radii for these galaxies. For 
large galaxies we assumed that the density asymptotically goes as 
$\sim r^{-4}$ (Jaffe 1987; White 1987; Tremaine 1987; Hjorth \& Madsen 1991) 
in the outer parts. Assuming a constant mass-to-light ratio with radius this 
means that the accumulated luminosity can be estimated as
\begin{equation}
L(R) = L_{\rm TOT}\left (1-{R_s\over R}\right)+\pi s R^2
\end{equation}
at large radii, where $s$ is the sky value. The value of the scale radius
$R_s$ depends on the luminosity profile at smaller radii. Its value is 
$0.67R_e$ for the model of Brainerd, Blandford \& Smail (1996) and 
$0.87R_e$ for the Hernquist (1990) model. The sky value $s$, the total 
luminosity $L_{\rm TOT}$, and the scale radius $R_s$ were determined from 
a fit to this equation in the outer parts of the profile ($R \gg R_e$; 
typically 4--6 $R_e$ in our case). Finally, the sky subtracted luminosity
inside an annulus of radius $R_e$ was determined directly from the image 
(central few arcsec) and fit. In addition, effective radii were also 
determined by fitting an \rq\ law to the surface-brightness profile or by 
fitting an \rq\ growth curve to the observed growth curve. The latter fits 
were kindly carried out by I. J\o rgensen to mimic the data analysis of JFK. 
For these fits the \rq\ growth curve was fitted in the range 43 arcsec 
(36 arcsec for the two smaller galaxies) to an outer radius at which the
surface brightness was 25 mag arcsec$^{-2}$ in $R_C$.

The errors in the derived effective parameters are entirely dominated by 
systematic effects arising from the particular assumptions that go into the 
procedures for their determination. For this reason formal fitting errors for 
effective parameters will not be quoted in what follows.

\subsection{Literature data}

For comparison with JFK, who used velocity dispersions and $\mg$ line
indices inside an equivalent circular aperture at Coma of 3\farcs4, we 
need to estimate spectroscopic parameters as they would appear
in a circular aperture of radius
$\sim16\arcsec$ (for a distance ratio of $\sim 9$). In principle this 
can be achieved by applying aperture corrections to central values,
e.g., from the catalog of McElroy (1995). 
Such a procedure was followed for the determination of $\mg$ line indices,
where we applied an aperture correction 
of $-0.045$ to central values (J\o rgensen \etal\ 1995).
However, to evaluate accurate velocity dispersions, which are
more critical to the FP analysis, we instead examined recent, high-quality, 
long-slit spectroscopic data from the literature, as described below.

The effective measured velocity dispersion, $\sigma$, inside a given circular 
aperture arises as the combined effect of pressure and bulk unresolved rotation.
Assuming that the true intrinsic velocity dispersion, $\tilde \sigma$, is 
isotropic, the observed velocity dispersion can be expressed as

%For a
%disk at inclination $i$ exhibiting pure circular motion with a (mean) 
%rotation velocity $V_c$ we have 
%\begin{equation}
%V_{\rm obs}=V_c \sin i \cos \theta
%\end{equation}
%(Mihalas \& Binney 1981).
%Here $V\equiv V_c \sin i$ is the observed velocity along
%the major axis, $\theta$ is the polar angle in the plane of the 
%galaxy which is related to the polar angle on the sky through
%$\tan \theta= \sec i \tan \phi$. Finally, the inclination may
%be estimated as $\cos i = b/a=1-e$. Averaging over $\phi$ we get
%\begin{equation}
%\left <V_{\rm obs}^2\right>=q V^2
%\end{equation}
%with
%\begin{equation}
%q=\left < \cos^2 \tan^{-1}(\tan \phi/(1-e))\right >.
%\end{equation}
%This parameter varies from 0.50 for a face-on disk to 0.67 for a galaxy
%with $e=0.5$. For the galaxies in question $q$ varies from
%0.53 (NGC 3379) to 0.65 (NGC 3377). In principle we should also
%include the effect of the roughly linear increase in the central
%parts, but this region is relatively small compared to the
%17\arcsec\ and it is easy to show that the relevant prefactor $q$ is roughly
%the same.

%Given the uncertainties
%in this approach we have estimated the observed velocity dispersion 
%including unresolved rotation as

\begin{equation}
\sigma^2=\tilde \sigma^2+qV^2.
\end{equation}
Here $V$ is the effective rotation velocity observed along the major axis. 
The velocity fields, structure, and inclination of E/S0 galaxies are
in general unknown, making $q$ uncertain. Instead we shall resort to a
naive relation,
\begin{equation}
V(\phi)=V \cos^n \phi,
\end{equation}
where $\phi$ is the angle on the sky relative to the major axis. From this
it immediately follows that
\begin{equation}
q=\left < \cos^{2n} \phi \right >.
\end{equation}
For an inclined disk one expects that $0<n<1$ which translates into $0.5<q<1$. 
In addition, the inner linear rise of the rotation curve and large 
ellipticities tend to lower this value. Estimating $q$ in a statistical sense 
using $\sigma^2$ as a measure of the amount of effective kinetic energy 
Prugniel \& Simien (1994) found $\left < q \right >=0.81$. J\o rgensen, 
Franx \& Kj\ae rgaard (1993) used $q=1$ in a similar study. For NGC 3379 the 
available spectroscopy (see Statler 1994) yields $n=2.5$ and $q=0.34$, but 
this galaxy in not strongly affected by rotation. Saglia, Bender \& Dressler  
(1993) gave an observational example for the edge-on spiral NGC 4697 which 
yields $q=0.87\pm0.63$. In this \paper\ we shall use $q=0.75\pm0.25$. The 
uncertainty in this quantity is treated as a random (as opposed to a 
systematic) uncertainty since $q$ is likely to be different for different 
galaxies, due to different viewing angles, ellipticities, and internal 
kinematics.

\subsection{Notes on individual galaxies}

The photometric and spectroscopic parameters for each galaxy
are discussed here and summarized in Table~1.

\placetable{tbl-1}

\begin{deluxetable}{ccccccccc}
%\small
%\footnotesize
%\scriptsize
%\tablewidth{7truecm}
%\tablenum{}
\tablecaption{Parameters for \leo\ group early type galaxies
\label{tbl-1}}

\tablehead{
\colhead{Galaxy} &
\colhead{$m_{R_C}$} &
\colhead{$R_e$}&
\colhead{$A_B$\tablenotemark{a}}&
\colhead{$\log \left < I_r\right >_e$\tablenotemark{b}} &
\colhead{$\tilde \sigma$}&
\colhead{$V$}&
\colhead{$\sigma$}&
\colhead{$\mg$} \nl
\colhead{(NGC)} &
\colhead{(mag)} &
\colhead{(arcsec)} &
\colhead{ (mag)} &
\colhead{($L_\odot$\,pc$^{-2}$)} &
\colhead{(${\rm km\,s}^{-1}$)} &
\colhead{(${\rm km\,s}^{-1}$)} &
\colhead{(${\rm km\,s}^{-1}$)} &
\colhead{}
}
\startdata
3377     &  9.40 & 56.9 & 0.07 & 2.37 & 97  & 75 & 117 & 0.242 \nl
3379     &  8.54 & 55.9 & 0.05 & 2.72 & 194 & 42 & 197 & 0.304 \nl
3384     &  9.32 & 29.4 & 0.07 & 2.98 & 140 & 80 & 156 & 0.265 \nl
3412     & 10.13 & 22.4 & 0.06 & 2.88 & 103 & 57 & 114 & 0.188 \nl
3489\tablenotemark{c} &  9.75 & 18.4 & 0.02 & 3.20 & 115 & 60 & 126 & 0.150 \nl
\enddata

\tablenotetext{a}{From RC3. The Cousins $R$ extinction is $A_{R_C}=
0.53A_B$ according to Cardelli, Clayton, \& Mathis (1989).}
\tablenotetext{b}{Calculated as 
$\log \left < I_r\right >_e=-0.4[m_{R_C}(R_e)+0.37-0.53A_B+ 2.5\log \pi +
5\log R_e - 26.40-10\log(1+z_\leo)-2.5\log(1+z_\leo)]$ which
transforms into Gunn $r$ and corrects for Galactic extinction,
cosmological surface-brightness dimming and spectral bandwidth 
(k-correction).}
\tablenotetext{c}{Not used in construction of the fundamental plane as outlined in text.}

\end{deluxetable}

{\bf NGC 3377}
Classification: E5$-$1, T$=-5$ (RC3); E6 (RSA).\\
NGC 3377 has a companion galaxy, NGC 3377A, which was masked out from 
the image.  Furthermore, the galaxy extends further out than the corners of 
the frame, so the sky-level and galaxy parameters were determined using 
Eq.~(1). Otherwise, the galaxy fitting was straightforward and 
we find $m_{R_C}=9.40$ and $R_e=56\farcs9$. From a direct fit to the \rq\ law
we find $R_s/R_e=0.77$ and $R_e=55\farcs8$.
%The ellipticity is taken to be $e=0.45$ (Scorza \& Bender 1995). 
%The galaxy is possibly in the background (Ciardullo \etal\ 1993) by 0.1 mag. 
The velocity data come from Bender, Saglia \& Gerhard (1994). Along the major 
axis inside $16\farcs83$ they find a fitted mean velocity dispersion of 
$85\pm10\kms$, a mean rotation velocity of $87.5\pm1.0\kms$, 
$\left< H_3\right> <-0.093$ and $\left< H_4\right>  = 0.048\pm 0.013$. From 
their Fig.~4 we find a maximum rotation velocity of $100\kms$ at $4\farcs5$,
and a constant rotation velocity at a level of $85\kms$ outside $6\arcsec$
(fitted values). The velocity dispersion drops from a central value of 
$150\kms$ to $100\kms$ at $6\arcsec$ and a mean of about $80\kms$ out to 
$34\arcsec$. These values indicate that the bulk of the velocity moments are 
contributed by the outer parts and are not strongly affected by the inner 
structure or outer aperture size. We adopt a value of $\left< H_3\right> =-0.1$ 
and find from their Fig.~3 that $\tilde \sigma/{\tilde \sigma_{\rm fit}}= 1.14$ 
and $(V-v_{\rm fit})/{\tilde \sigma_{\rm fit}}=-0.15$. We thus arrive at
$\tilde \sigma=97\pm 11\kms$ and $V= 75\pm 2\kms$, and hence 
$\sigma=117\pm12\kms$. The aperture corrected line index $\mg=0.242\pm0.014$
(J\o rgensen \etal\ 1995). For comparison, the same data yields an aperture 
corrected central velocity dispersion of $\sigma=135\kms$.

{\bf NGC 3379}
Classification: E1, T$=-5$ (RC3); E0 (RSA).\\
NGC 3379 is a `standard E galaxy' discussed in detail by Capaccioli 
\etal\ (1990) and Statler (1994). In our analysis, the neighboring galaxy 
NGC 3384 was subtracted iteratively by masking and fitting NGC 3384 and 
NGC 3379 interchangeably several times. The light profile of NGC 3379 does not 
reach the sky level either so the comment made for NGC 3377 also applies here. 
We find $m_{R_C}=8.54$ and $R_e=55\farcs9$ in agreement with Capaccioli \etal\ 
(1990) and Sodemann \& Thomsen (1994). The direct fit to the \rq\ law yields
$R_s/R_e=0.73$ and $R_e=53\farcs 4$. The galaxy is a minor axis rotator 
(see Statler 1994). We again take the velocity data from Bender \etal\ (1994) 
who report values of fitted mean velocity dispersion of $180\pm10\kms$, mean 
rotation velocity of $50.8\pm4.9\kms$, $\left< H_3\right> =-0.029\pm 0.028$ and 
$\left< H_4\right>  = 0.041\pm 0.028$ along the major axis inside a radius of 
$17\farcs63$. The velocity dispersion falls off less sharply than in NGC 3377, 
whereas the rotation curve increases steadily. We adopt corrections for the 
velocity profile distributions of  $\tilde \sigma/{\tilde \sigma_{\rm fit}}
=1.08\pm0.05$ and $(V-v_{\rm fit})/{\tilde \sigma_{\rm fit}}=-0.05\pm0.05$ and
arrive at $\tilde \sigma=194\pm 14\kms$ and $V= 42\pm 11\kms$, and hence 
$\sigma=197\pm14\kms$. The corrected $\mg=0.304\pm0.014$ (J\o rgensen \etal\ 
1995). The same data yields an aperture corrected 
central velocity dispersion of $\sigma=200\kms$.

{\bf NGC 3384}
Classification: SB(s)0$^-$:, T=$-3$ (RC3); SB0$_1$(5) (RSA).\\
Contaminating light from NGC 3379 was removed as described above. The fit 
shows that NGC 3384 is much less extended and the sky value could be 
determined from the corners of the frame. We find $m_{R_C}=9.32$ and 
$R_e=29\farcs4$ in reasonable agreement with the value of $R_e=22\farcs5$ 
found in the $B$ band by Busarello \etal\ (1996). The central velocity 
dispersion along the minor and major axes is $160\pm5\kms$ but then drops 
to about $100\kms$ at a radius of 16\arcsec\ (Busarello \etal\ 1996; 
Fisher 1996). We estimate the mean value inside 16\arcsec\ to be 
$\tilde \sigma=140\pm10\kms$, in agreement with the value of  
$\tilde \sigma=138\pm10\kms$ reported by Fisher, Franx, \& Illingworth (1996).
% and slightly smaller than
%that found from McElcroy's (1995) catalog
%of central velocity dispersions corrected as described ($\sigma=156\kms$).
The rotation around the minor axis is about $80\pm 10\kms$ and there is no 
rotation around the major axis (Busarello \etal\ 1996; Fisher 
1996), and hence we obtain $\sigma=156\pm11\kms$.
%A slightly higher rotation velocity
%%around the minor axis
%was measured by Tremblay \etal\ (1995) 
%using planetary nebulae to be $125\kms$ (the velocity dispersion they found 
%with considerably less spatial resolution was $100\kms$, consistent with 
%the stellar value outside 10\arcsec).
The corrected $\mg$ from Fisher \etal\ (1996) is 0.281 but this should 
be compared with 0.249 from Worthey, Faber, \& Gonzalez (1992).
We adopt $\mg=0.265\pm0.016$.

{\bf NGC 3412}
Classification: SB(s)0$^0$, T$=-2$ (RC3); SB0$_{1/2}$(5) (RSA).\\
The galaxy fit was straightforward and the sky value could be determined from 
the corners of the frame. We find $m_{R_C}=10.13$ and $R_e=22\farcs4$. The 
velocity dispersion reported by Fisher \etal\ (1996) is 
$\tilde \sigma=103\kms$ with an estimated uncertainty of $10 \kms$.
%consistent with the value from McElroy (1995) corrected to give
%$\sigma=97\kms$. 
There is no rotation around the minor axis whereas the rotation around the 
major axis shows a steady increase out to 27\arcsec\ (Fisher 1996).
We adopt the rotation velocity at 10\arcsec\ of $57\kms$ and assign a large 
uncertainty of $20\kms$ to this value. Thus we estimate $\sigma=114\pm12\kms$.
The corrected $\mg=0.188$ (Fisher \etal\ 1996).

{\bf NGC 3489}
Classification: SBA(rs)0$^+$, T$=-1$ (RC3); S0$_3$/Sa (RSA).\\
This galaxy is spiral like and has hot gas, dust, etc., with line emission 
seen in its nuclear spectrum (Ho, Filippenko, \& Sargent 1995). The galaxy 
fit was straightforward and the sky value could be determined from the corners 
of the frame. We find $m_{R_C}=9.75$ and $R_e=18\farcs4$. The velocity data 
are taken from Bertola \etal\ (1995) who obtained spectra along the major 
axis. The velocity dispersion is roughly constant out to $20\arcsec$ at a 
mean level of $\tilde \sigma=115\pm10\kms$ and the rotation rate is nearly 
constant outside 2\arcsec\ at a mean value of $V=60\pm10\kms$. Thus we 
estimate $\sigma=126\pm10\kms$. The corrected $\mg$ line index is 
%calculated from (Mg)$_0=0.085$
%(Burstein 1979) as Mg$_2=1.33$(Mg)$_0+0.086-0.045$ (Faber \etal\ 1985).
%Can also be obtained from 
%$0.195-0.045=0.150$ (Worthey \etal\ 1992).
$0.150$ (Worthey \etal\ 1992). The galaxy has several `boxy' emission lines 
due to multiple velocity components in the nuclear region (Ho \etal\ 1995). 
This complicated velocity structure may indicate that $\sigma$ is 
underestimated for this galaxy.

\section{THE FUNDAMENTAL PLANE IN LEO-I}

Figure~2 shows the $\mg$ values as a function of $\log \sigma$. Also plotted 
is the mean $\mg$--$\log \sigma$ relation for the JFK sample. The larger 
galaxies, NGC 3377, NGC 3379, and NGC 3384 are consistent with this relation 
and can be used with confidence for the FP. It is clear, however, that the 
two small late-type lenticulars, especially NGC 3489, fall below the line. 
As discussed by JFK the scatter in the FP can be significantly reduced 
and artifacts avoided if such galaxies are excluded from the fit. NGC 3489 
also has by far the highest mean surface brightness inside the effective 
radius, possibly due to recent star formation: NGC 3489 is the only
galaxy with clear emission lines, whereas the other four galaxies have
genuine early-type spectra (Ho \etal\ 1995). 
%Galaxies with high
%surface brightness generally induce deviations from the FP
%(Gregg 1995). 
Finally, NGC 3489 has the largest projected distance from the group center 
and so this galaxy (if any) is the most likely to be either in the foreground
or background relative to the rest of the \leo\ galaxies. For these reasons
we exclude NGC 3489 in the fits, but include it in the plots for completeness.

The fundamental plane that we fit is
\begin{equation}
\log R_e = \alpha \log \sigma +
\beta \log \left < I_r \right >_e  + \gamma,
\end{equation}
where $R_e$ is measured in arcsec, $\sigma$ in$\kms$, 
$ \left < I_r \right >_e$ is the
mean intensity inside the effective radius in Gunn $r$ measured 
in ${\rm L_\odot\,pc}^{-2}$, and $\gamma$ is the distance-dependent
zero point of the relation. [It is straightforward to correct from 
Cousins $R_C$ to Gunn $r$ for E galaxies. We adopt $r-R_C=0.37$ 
(J\o rgensen 1994)]. The global values adopted by JFK are $\alpha=1.24$ and 
$\beta=-0.82$. In principle FP distances may be affected by the observed 
magnitude range, morphological makeup of the sample (E/S0 ratio) and 
environment.  According to JFK, their version of the FP excludes biases 
above 1\% due to magnitude and morphological selection. The situation with 
respect to environment is more uncertain and is discussed below.

%The FP fit is performed by minimizing the
%mean absolute deviation perpendicular to the FP, assigning weights to
%each point inversely proportional to this uncertainty squared. We only
%fit the FP to the four brighter galaxies since NGC 3489 clearly
%falls below the line defined by the other galaxies, probably
%for the reasons discussed above. 

Figure~3 shows the FP for the \leo\ galaxies with the slopes $\alpha$ and 
$\beta$ fixed at their global values. The FP offset is computed as the median 
offset (JFK) of the four larger galaxies, the slope being consistent with 
that found by JFK. NGC 3489 clearly falls below the line defined by the other 
galaxies, probably for the reasons discussed above. We obtain 
$\gamma_{\leo} = 1.150$. Other ways of computing the offset, including mean 
and weighted mean give same result within $\pm0.007$. The observed formal 
rms scatter in $\gamma$ is 0.024 or 6 \% in $R_e$. It is interesting to note 
that this is significantly smaller than expected from the 14 \% intrinsic 
scatter plus observational errors found by JFK.
%The observational uncertainty in the zero point of $(0.022^2/4+0.005^2)^{1/2}=
%0.014$ or 3 \% in distance.

%that expected.  The total uncertainty (observational and
%expected intrinsic scatter added in quadrature) in this value is 
%9 \%: 
%The data points for which the effective radii of the galaxies are uncertain 
%would essentially move parallel to 
%the FP (JFK) and so this uncertainty only contributes an error of
%about 2\% (JFK).
%The uncertainties in the effective velocity dispersions, arising
%partly from the uncertain correction due to unresolved rotation,
%thus entirely dominate the FP fit.
%The uncertainty attributed to each point is obtained by
%adding the uncertainty in $\log \sigma$, 2 \% from the
%uncertainty in $R_e$ and the expected 13 \% intrinsic
%scatter in quadrature. 
%
%{\bf [JH--Uncertainties!?]}.

%This may indicate that the intrinsic scatter of 13 \% 
%may be overestimated in \leo. In fact, the observed scatter is smaller
%than that expected from the estimated observational errors alone and
%is consistent with there being no intrinsic scatter in the \leo\ FP.
%However, the estimated  13 \% intrinsic scatter also takes systematic
%zero-point differences from cluster to cluster into account and so the
%observed scatter around the \leo\ FP is not necessarily
%inconsistent with this value.

Another approach would be to consider the zero-point obtained from NGC 3379 
alone. The motivation for this is that NGC 3379 is a bona fide elliptical
galaxy, with normal colors, stellar populations (as witnessed e.g.\ by the 
$\mg$ index), small ellipticity, no evidence for an embedded disk 
(Statler 1994), and a small rotation velocity.
Thus it escapes the inherent 
uncertainty present in the prescription for correction of unresolved rotation 
(eq.~2) and other possible systematic errors that could arise from using the 
other, later-type,
\leo\ galaxies. In view of the very small intrinsic scatter observed in 
the FP for the central regions of Coma alone 
($<5 \%$) (J\o rgensen \etal\ 1993)
it is plausible that the intrinsic uncertainty for NGC 3379 
alone is significantly smaller than the formal 14 \% for the global FP. The 
zero point derived for NGC 3379 only is 1.135 with an estimated observational 
uncertainty of 0.030.

The final possibility is to treat the departure from the  $\mg$--$\log \sigma$
relations as an extra parameter. Environment may be important given that 
\leo\ is a relatively poor group, comparable to the smallest groups studied by 
JFK. Guzm\'an \etal\ (1992) found a correlation of the $\mg$--$\log \sigma$ 
relation with radial distance from the center of the Coma cluster, and argued 
that this was consistent with galaxies in less dense environments having on 
average younger stellar populations. This allowed Guzm\'an and Lucey (1993) to 
formulate an `environment-independent' FP which incorporated an $\mg$ term,
evaluated using galaxy evolution models, to 
correct for this systematic difference. JFK also identify similar behavior for 
the $\mg$--$\log \sigma$ relation, although they only find a marginal 
environmental effect of $3\pm3$ \% for distances obtained with their `classic' 
FP over the full range of groups and clusters in their study. 
Including an empirically determined
$\mg$ term in the FP should in any case provide some robustness to 
differences in stellar populations between galaxies, which may in turn vary
systematically with environment, whether such differences are 
due to variations in
age or metallicity.  Interestingly JFK found that, while the 
observed scatter in the FP was not significantly reduced with the
extra term, it did in some cases 
bring outliers with discrepant $\mg$ line indices closer to the global FP. 
For this $\mg$ FP $\alpha=1.05$, $\beta=-0.78$ and 
$-0.40 \Delta\mg$ is added to the rhs of the FP equation (5). The significance 
of this term is at the 1.5--2 sigma level (JFK), and it 
is evident that it
brings NGC 3489 much closer to the FP defined by the other four galaxies. 
For these we find $\gamma(\mg)_{\rm \leo}=1.449\pm0.014$ and an observed 
formal scatter of 6 \%. 

We note that, if instead we had used the effective radii and intensities as
determined from the fit of \rq\ growth curves, the corresponding numbers are 
$\gamma_{\leo}=1.144\pm0.005$, $\gamma_{\rm NGC 3379}= 1.148\pm0.030$, and 
$\gamma(\mg)_{\rm \leo}=1.445\pm 0.015$. It is reassuring that the zero points 
remain unchanged within the uncertainties.

{From} $\gamma_{\rm Coma}=0.182\pm0.009$ (JFK) and 
$\gamma(\mg)_{\rm Coma}=0.497\pm0.009$
(I. J\o rgensen, private communication) we find zero-point differences 
between \leo\ and Coma of $0.968\pm0.015$ (FP), $0.953\pm0.031$ (NGC 3379), 
and $0.952\pm0.017$ ($\mg$-FP). (The respective numbers from growth curve 
fitting are  $0.962\pm0.010$ (FP), $0.966\pm0.031$ (NGC 3379), and 
$0.948\pm0.017$). 
Here the uncertainties are estimated from the observed scatter
around the \leo\ FP, and clearly
the different values are consistent within these errors.
We adopt a value of $0.96\pm0.03$ where we now assume a conservative
uncertainty of $0.057/\sqrt{4}$ which is estimated from the expected intrinsic 
scatter of 14 \% in distance per galaxy. This corresponds to 7 \% in the 
distance ratio. It should be noted that this uncertainty 
is much higher than the 
estimated individual uncertainties and makes the derived distance ratio a 
robust value. We thus derive the angular diameter distance ratio of 
$9.12\pm0.64$ or a luminosity distance ratio of $9.51\pm0.67$ between Coma and 
\leo. 

Tanvir \etal\ (1995) obtained a distance to the \leo\ group of 
$11.6\pm0.9$ Mpc 
based on HST observations of Cepheids in the spiral galaxy M96. 
Using the revised Cepheid period--luminosity calibration
of Tanvir (1997) and incorporating the so-called
``long-exposure correction'' (Hill \etal\ 1997),
this becomes $11.3\pm0.9$ Mpc.
The distances 
to the core of \leo\ and M96 are tied together 
in a unique way, due to the giant
200 kpc diameter H\,{\sc i} ring (Schneider 1989) which orbits the central 
galaxies and interacts with M96.  This distance is supported by the recent
determination of the distance to NGC 3379 of $11.4\pm 0.3$ Mpc from SBF 
measurements (Sodemann \& Thomsen 1996) (adopting a distance to M32 of 0.77 
Mpc). Sakai \etal\ (1997) report a distance of $11.5\pm 1.6$ Mpc from 
detection of the TRGB 
using deep HST data. It should be mentioned, however, that 
several other recent estimates of distances to \leo\ galaxies indicate a 
slightly lower distance (see Yasuda \& Okamura 1996), e.g., $10.05\pm0.88$ Mpc 
from Cepheids in M95 (Graham \etal\ 1996, although this reduces
further to about 9.6 Mpc if the Tanvir (1997) Cepheid calibration is
used), 10.7 Mpc to \leo\ from SBF 
(Tonry \etal\ 1997) and values around 10 Mpc from PNLF (Ciardullo \etal\ 1993; 
Feldmeir \etal\ 1997). For a \leo\ distance of $11.3\pm0.9$ Mpc 
the derived distance ratio between \leo\ and Coma implies a Coma distance of 
$108\pm 12$ Mpc.  
This result is consistent with the GCLF distance to the Coma 
galaxy NGC 4881 of $108^{+\infty}_{-11}$ Mpc found by Baum \etal\ (1995) and
the SBF distance to the same galaxy of Thomsen, Baum, \& Hammergren  (1997). 

Using the JFK FP exponents, the \leo\ FP, the $\mg$-$\sigma$ relation (see 
Figure~2) and the distance to \leo\ we can calibrate the FP as an angular 
diameter distance indicator in terms of the observed effective radius 
$R_e$ (in arcsec), the effective intensity in Gunn $r$ $\left < I_r\right>_e$ 
(${\rm L_\odot\,pc}^{-2}$), the velocity dispersion $\sigma$ (km s$^{-1}$), 
and the $\mg$ index as
\begin{equation}
D=10^{-\log R_e+a\log \sigma 
+ b\log \left< I_r\right>_e+c\mg +d}\ \ {\rm Mpc}
\end{equation}
with $(a,b,c,d)=(1.24,-0.82,0,2.194)$ for the `classic' FP (5) 
or $(1.128,-0.78,-0.4,2.447)$ for the $\mg$ FP. The `predicted' values for the 
four FP calibrator galaxies are 11.7 Mpc, 11.3 Mpc, 10.3 Mpc, 11.6 Mpc (where 
we have taken the mean of the two derived distances) indicating the usefulness 
of this distance indicator.

\section{THE HUBBLE CONSTANT}

For the recession velocity of Coma we use $7200\kms$ (Faber \etal\ 1989)
%(Lynden-Bell \etal\ 1988) 
and adopt a conservative estimate of the possible peculiar velocity of 
$\pm400 \kms$ (Han \& Mould 1992; Bahcall \& Oh 1996) which also
allows for the intrinsic substructure in the Coma cluster 
(Colless \& Dunn 1996). In the terminology of Jerjen and Tammann (1993)
our result then implies a ``cosmic velocity'', of \leo\ of 
$757\pm68\kms$. Adding the uncertainties in quadrature we arrive at an 
uncertainty of 12 \% in the Hubble constant: thus we get 
$H_0=67\pm8\ho$ fixing to Coma.

Alternatively, tying \leo\ to the larger frame of all 11 clusters in the 
sample of JFK, using the observed redshifts in the CMB, yields a mean or 
median value of about $800\kms$ with an rms scatter of $65\kms$. This 
value is essentially
unchanged if we the exclude the small nearby group Doradus and the 
peculiar cluster S639 (JFK) but the scatter is reduced to $40\kms$.
% or a formal uncertainty of $13 \kms$. 
Relative to this larger cluster frame,
which has a median redshift of $\sim5000\kms$, we then find a mean value of 
$796\pm57\kms$ and a Hubble constant of $H_0=70\pm7\ho$. We have not corrected 
here for the effects of Malmquist or incompleteness biases, however these are 
expected to be small ($<1.5$ \%) given the numbers of galaxies observed in 
each cluster (JFK). 
Reassuringly, this ``cosmic velocity'' is fully consistent with that
found for the Leo-I group by the ``7 Samurai'' of $783\pm126\kms$ based
on the $D_n$--$\sigma$ relation (Faber \etal\ 1989). 
The higher value of $H_0$ obtained by Tonry \etal\ (1997)
using the SBF method to zero-point the $D_n$--$\sigma$ relation
is explained in part by the calibration: they find
a distance of 10.7 Mpc for the \leo\ group. The remaining difference is
nevertheless marginally significant given the quoted errors. This may 
reflect our smaller sample size (only four galaxies used
in the calibration) or difficulties in correcting their sample for
selection biases.

%The derived cosmic velocity is consistent with that of 7S 
%who found $783\pm???\kms$. Adopting some kind of average of the 
%recent \leo\ distance estimates of say $10.7\pm1$ Mpc (see \S 3), then
%tying to the JFK frame yields $H_0=74\pm9\ho$, consistent with most
%recent estimates of the Hubble constant
%(Dahle, Maddox, \& Lilje 1994; 
%Baum \etal 1995; 
%Grogin \& Narayan 1996; 
%Corbett, Browne, \& Wilkinson 1996; 
%Ruiz-Lapuente 1996; 
%Riess, Press, \& Kirshner 1996; 
%Giovanelli 1996; 
%Kundic \etal\ 1996;
%Tonry \etal\ 1997; 
%Thomsen \etal\ 1997).

%Our determination of $H_0$ is conservative.
%The quoted uncertainty in $H_0$ can however be brought down by introducing
%additional data and assumptions. For example, if we adopt a distance
%to NGC 3379 of $11.5\pm0.5$ Mpc consistent with Cepheid,
%SBF, and TRGB measurements, an intrinsic scatter for
%NGC 3379 of 5 \% as in the Coma cluster, and 
%a 2 \% uncertainty in the
%recession velocity of the Coma cluster (Mould 1996) we arrive at an 
%uncertainty of only 7.7 \% or $H_0=65\pm5\ho$.

There is plenty of room for improvement in this approach. Wide-field CCD 
frames should map the distribution of light in the larger galaxies to get more 
precise values of the photometric quantities. However, the largest uncertainty 
in the FP parameters are the effective velocity dispersions. Long-slit spectra 
taken in several positions over the galaxy should eliminate this uncertainty. 
The intrinsic uncertainty in the FP for individual galaxies, including 
environmental effects, should be better understood, possibly by including
more information on the stellar population, e.g.\ the H$\beta$ line index. 
Indeed, the most important worry in using the FP as a distance indicator is 
possible environmental effects. For example, if galaxies in low-density 
environments like \leo\ have on average younger stellar populations than the 
dense core of Coma (as discussed by Guzm\'an and Lucey 1993 and JFK), then
our distance to Coma obtained from the `classic' FP would be an overestimate, 
although, as we have seen, the use of the $\mg$--FP suggests that 
environmental effects are not large.

It would also be interesting to calibrate the FP in the infrared for several 
clusters where the scatter may be smaller (Pahre, Djorgovski, \& de 
Carvalho 1995) and the systematic effects of variations in stellar populations
less (Guzm\'an and Lucey 1993). The uncertainty arising from the Coma peculiar 
velocity should be eliminated by establishing the FP at larger redshifts, 
$0.05<z<0.1$. On the theoretical side the mechanism behind the FP does not 
seem to be fully understood, although the FP does seem to work as a purely 
empirical relationship.  After all, the assumption of a `universal' FP
underlies the distance indicator used in this \paper, and although
there is observational support for this, a firm theoretical basis for
this is clearly needed (Renzini \& Ciotti 1993; Hjorth \& Madsen 1995; 
Capelato,
de Carvalho, \& Carlberg 1995; Ciotti, Lanzoni, \& Renzini 1996).

%Regarding the \leo\ group itself as a calibrator of the FP,
%upcoming ASCA observations are expected to provide additional direct evidence
%for the association of NGC 3379 and M96.
%Also, TRGB, SBF, and PNLF distances
%to NGC 3379 and other galaxies in the group are worth exploring in
%more detail because of its unique nature.

\section{CONCLUSIONS}

The main result of this \paper\ is the construction of the fundamental plane 
(FP) of the \leo\ group. The small observed scatter (6 \%) about this 
relation means that the result is insensitive to the particular choice of 
and weighting of the galaxy sample.

We have adopted the form of FP determined by JFK and hence find the relative 
distance between Coma and \leo\ to be $9.5\pm0.7$. In addition, using the 
distance to M96 in \leo\ ($11.3\pm0.9$ Mpc) yields a direct and accurate 
measurement of the Coma distance of $108\pm12$ Mpc. Finally, using the 
recession velocity of Coma ($7200 \pm400\kms$), we estimate the 
Hubble constant to be $H_0=67\pm8\ho$, or alternatively, if we consider 
the full cluster sample of JFK, with a median redshift $\sim5000\kms$, we find 
$H_0=70\pm7\ho$.

%The data seem to rule out both the short  ($H_0\ga 80\ho$) (Jacoby \etal\
%1992; Freedman \etal\ 1994; Yasuda, Fukugita, \& Okamura 1996) and long
%($H_0\approx 55\ho$) distance scales but
%within the estimated errors the results are consistent with the upper value 
%of $H_0$ of the long distance scale
%(60 \cf Sandage \etal\ 1996; Schaeffer 1996; Tammann \etal\ 1996) and the 
%lower value of $H_0$ of the short distance scale (Freedman \etal\ 1994). 
%Gravitational lensing
%(Dahle, Maddox, \& Lilje 1994; Grogin \& Narayan 1996) and SNIa 
%data with independent
%Cepheid distances also support this value (Riess, Press, \& Kirshner 1996;
%Ruiz-Lapuente 1996).

%The data are inconsistent with current SZ 
%estimates (well.., perhaps not) but/and consistent with cosmological values 
%inferred from gravitational lenses (Dahle, Maddox, \& Lilje 1994;
%Grogin \& Narayan 1996; Nair 1996). 
%All HST + Coma values agree on 
%a low $H_0 < 70$. SNIa?

%It is interesting to note that 
%existing observational 
%data are consistent with a simple cosmological model with 
%$H_0=65$, $\Omega=0.3$, $t_0=14$, $\Lambda=0$.

\acknowledgments

We are grateful to Inger J\o rgensen, David Fisher, Bjarne Thomsen, 
Raphael Guzm\'an, and Alfonso Arag\'on-Salamanca for help, communication of 
unpublished results, and many useful comments and discussions. JH acknowledges 
support from the Danish Natural Science Research Council (SNF).

\newpage
\begin{figure}
\vspace{17truecm}
\includegraphics{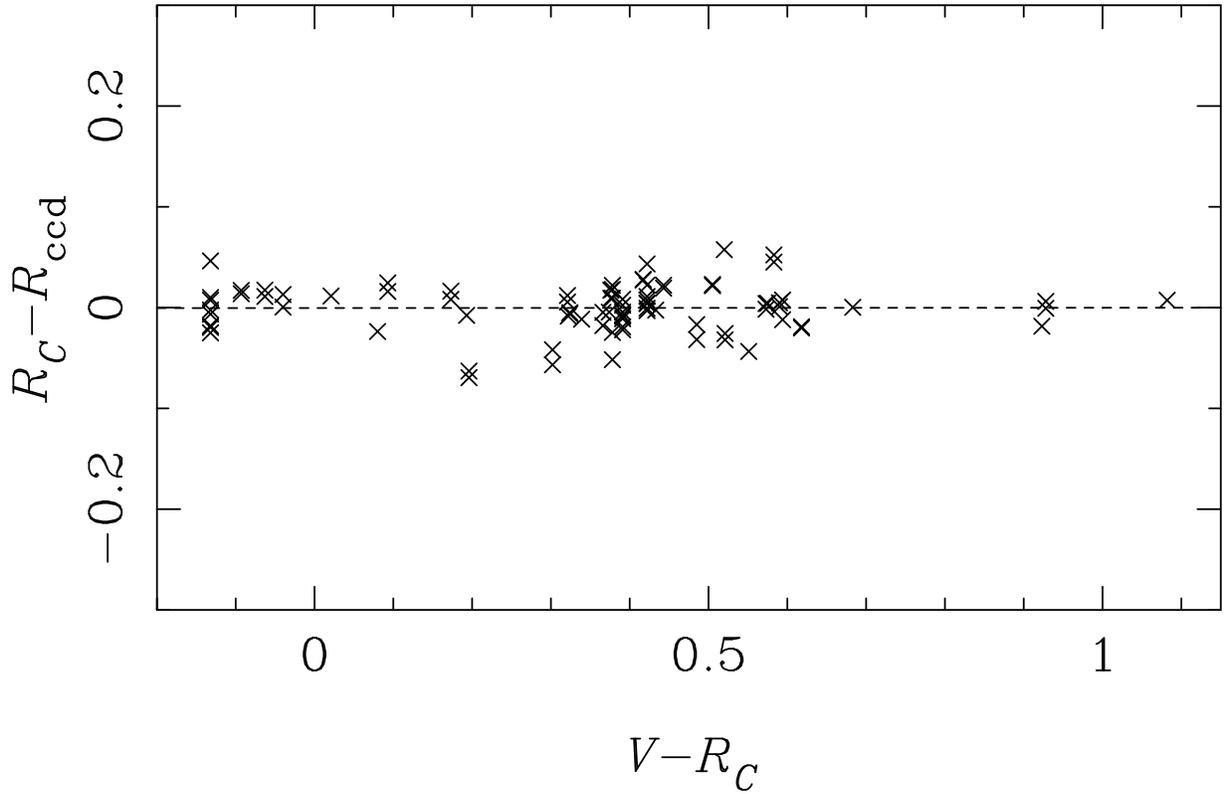}
\caption{
Comparison of standard star $R_C$ magnitudes and  measured 
instrumental magnitudes as a function of $V-R_C$ color.
The dashed line is a fit to the data.
The rms scatter around the line is only 0.02 mags and
there is no measurable color term or non-linearity.
\label{figure1}}
\end{figure}

\begin{figure}
\vspace{17truecm}
\includegraphics{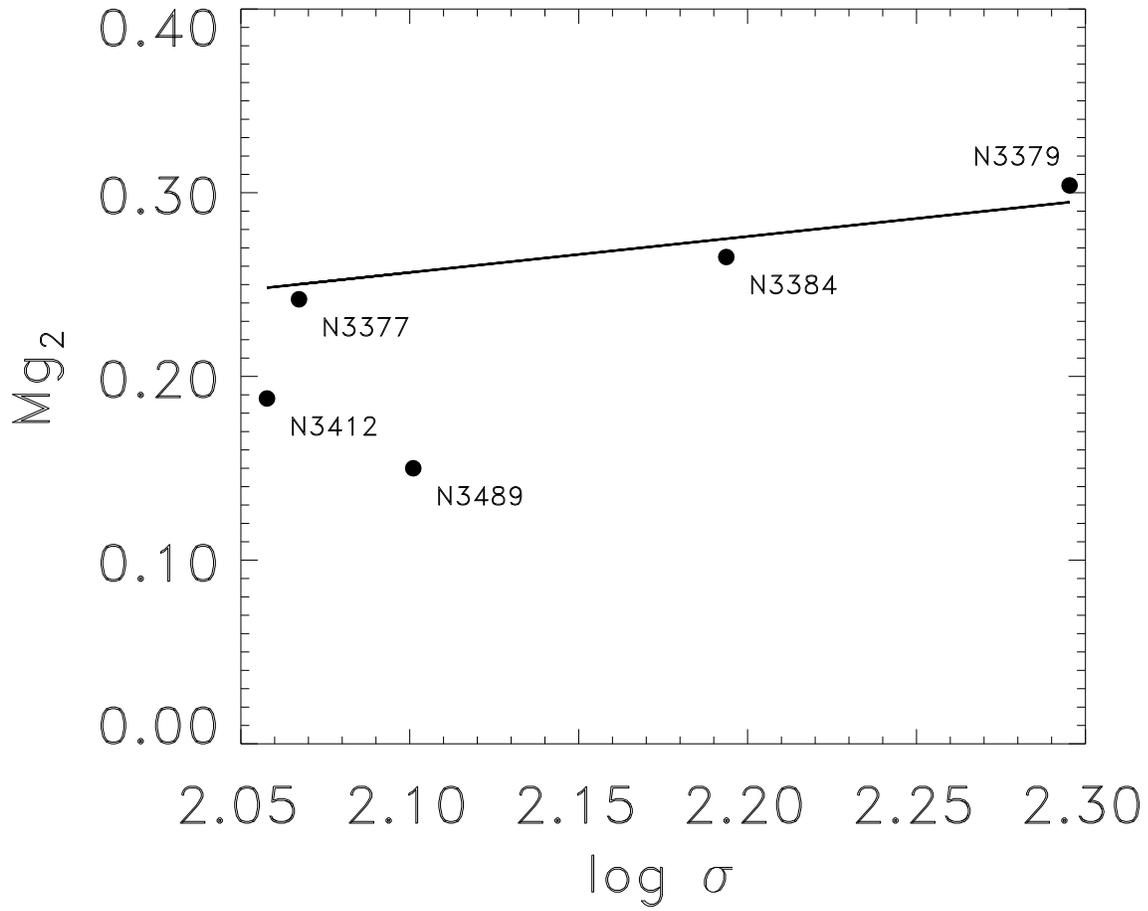}
\caption{
The $\log \sigma$--$\mg$ relation. The filled circles are \leo\
galaxies, the solid line is the $\log \sigma$--$\mg$ relation of JFK 
($\mg=0.196 \log \sigma -0.155$) and has
not been fitted to the data.
\label{figure2}}
\end{figure}

\begin{figure}
\vspace{17truecm}
\includegraphics{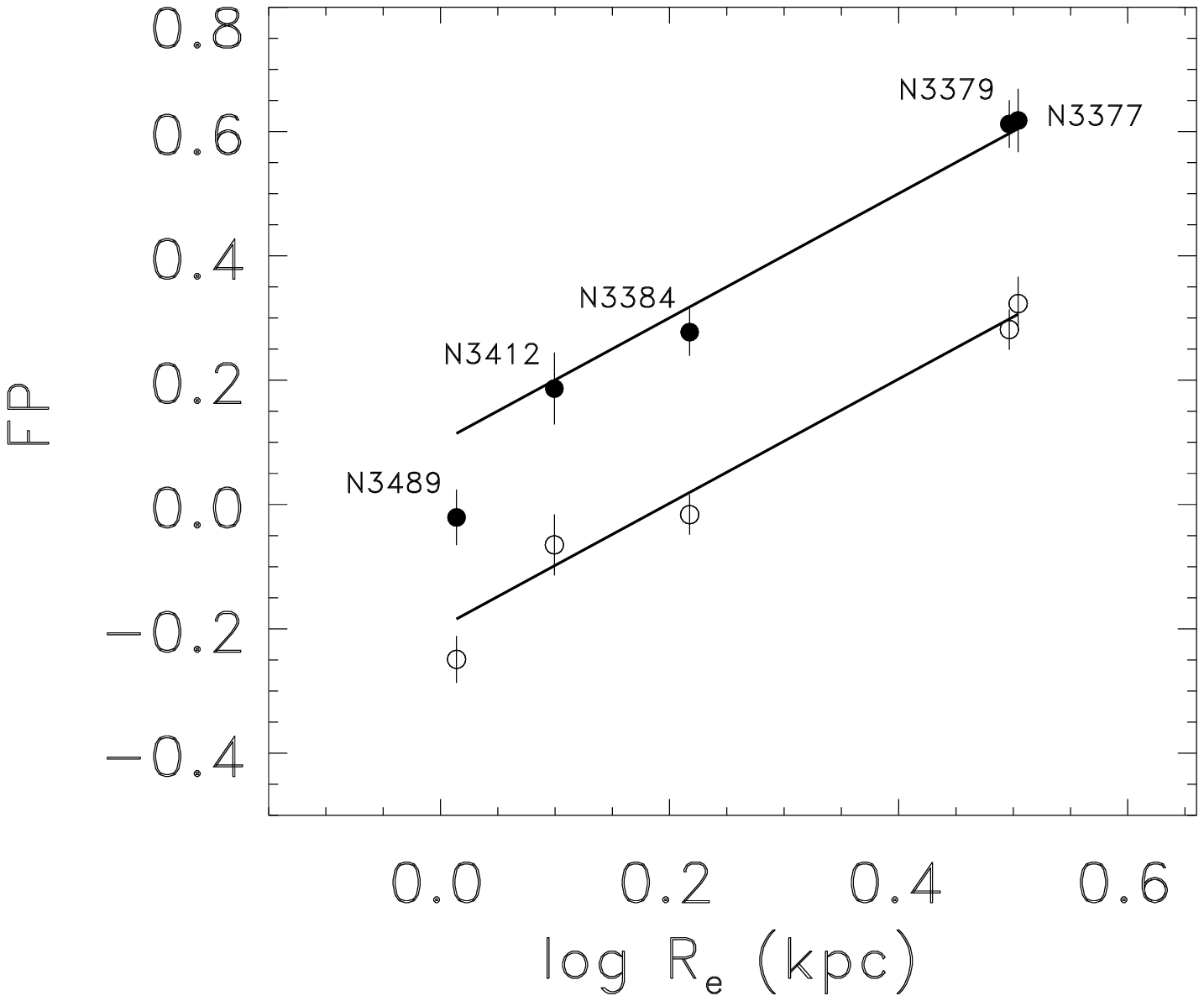}
\caption{
The fundamental plane in \leo.  The filled circles are `classic'
FP = $ 1.24 \log \sigma - 0.82 \log \left < I_r \right >_e$
values for \leo\ galaxies, the solid line through the points
is the JFK FP for the four larger galaxies. 
The open circles are $\mg$-FP =
$ 1.05 \log \sigma - 0.78 \log \left < I_r \right >_e
-0.40 \Delta\mg$ values for \leo\ galaxies and the
solid line through these points is the JFK $\mg$--FP for the four 
larger galaxies. 
\label{figure3}}
\end{figure}


\clearpage

\begin{references}
\reference{}
Bahcall, N. A., \& Oh, S. P., \apj, 462, L49

\reference{}
Baum, W. A. \etal\ 1995, \aj, 110, 2537

\reference{}
Bender, R., Saglia, R. P., \& Gerhard, O. E. 1994, \mnras, 269, 785

\reference{}
Bertola, F., Cinzano, P., Corsini, E. M., Rix, H.-W., \& Zeilinger, W. W.
1995, \apj, 448, L13

\reference{}
Brainerd, T. S., Blandford, R. D., \& Smail, I. 1996, \apj, 466, 623

%\reference{}
%Burstein, D. 1979, \apj, 232, 74

%\reference{}
%Burstein, D., \& Heiles, C. 1984, \apjs, 54, 33

\reference{}
Busarello, G., Capaccioli, M., D'Onofrio, M., Longo, G., Richter, G., \&
Zaggia, S. 1996, \aap, 314, 32

\reference{}
Capaccioli, M., Held, E. V., Lorenz, H., \& Vietri, M. 1990, \aj, 99, 1813

\reference{}
Capelato, H. V., de Carvalho, R. R., \& Carlberg, R. G. 1995, \apj, 451, 525

\reference{}
Cardelli, J. A., Clayton, G. C., \& Mathis, J. S. 1989, \apj, 345, 245

\reference{}
Ciardullo, R., Jacoby, G. H., \& Ford, H. C. 1989, \apj, 344, 715

\reference{}
Ciardullo, R., Jacoby, G. H., \& Tonry, J. L. 1993, \apj, 419, 479

\reference{}
Ciotti, L., Lanzoni, B., \& Renzini, A. 1996, \mnras, 282, 1

\reference{}
Colless, M., \& Dunn, A. M. 1996, \apj, 458, 435

%\reference{}
%Corbett, E. A., Browne, I. W. A., \& Wilkinson, P. N. 1996, in 
%Astrophysical Applications of Gravitational Lensing,
%IAU Symp.~173, eds. C.S. Kochanek \& J.N. Hewitt (Kluwer Dordrecht), p.~37

%\reference{}
%Dahle, H., Maddox, S. J., \& Lilje, P. B. 1994, \apj, 435, L79

\reference{}
de Vaucouleurs, G., de Vaucouleurs, A., Corwin, H. G., Buta, R. J.,
Paturel, G., \& Fouqu\'e, P. 1991, Third Reference Catalogue of
Bright Galaxies (Springer, New York) (RC3)

\reference{}
Djorgovski, S., \& Davis, M. 1987, \apj, 313, 59

\reference{}
Dressler, A. \etal\ 1987, \apj, 313, 42

%\reference{}
Faber, S. M. \etal\ 1989, \apjs, 69, 763

%\reference{}
%Faber, S. M., Friel, E. D., Burstein, D., \& Gaskell, C. M. 1985,
%\apjs, 57, 711

\reference{}
Feldmeir, J. J., Ciardullo, R., \& Jacoby, G. H. 1997, \apj, submitted

\reference{}
Ferguson, H. C., \& Sandage, A. 1990, \aj, 100, 1

\reference{}
Fisher, D. 1996, \mnras, submitted

\reference{}
Fisher, D., Franx, M., \& Illingworth, G. 1996, \apj, 459, 110

%\reference{}
%Freedman, W. L. et al.\ 1994, Nat, 371, 757

%\reference{}
%Freedman, W. L., \& Madore, B. F. 1996, in
%Clusters, Lensing, and the Future of the Universe, ASP Conf.\ Ser.\ 88,
%eds.\ V. Trimble, \& A. Reisenegger (ASP, San Francisco), p.\ 9

\reference{}
Garcia, A. M. 1993, \aaps, 100, 47

\reference{}
Graham, J. A. \etal\ 1997, \apj, in press
%John A. Graham
%Randy L. Phelps \& Wendy L. Freedman
%Abhijit Saha \& Laura Ferrarese
%Peter B. Stetson
%Barry F. Madore, N. A. Silbermann \& Shoko Sakai
%Robert C. Kennicutt, Paul Harding, Fabio Bresolin \& Anne Turner
%Jeremy R. Mould \& Daya M. Rawson
%Holland C. Ford
%John G. Hoessel \& Mingsheng Han
%John P. Huchra \& Lucas M. Macri
%Shaun M. Hughes
%Garth D. Illingworth \& Daniel D. Kelson

%Giovanelli, R. 1997, in 
%The Extragalactic Distance Scale,
%eds.\ M. Livio, M. Donahue, \& N. Panagia (CUP), in press

%\reference {}
%Gregg, M. D. 1995, \apj, 443, 527

%\reference {}
%Grogin, N. A., \& Narayan, R. 1996, \apj, 464, 92

\reference{}
Guzm\'an, R., Lucey, J., Carter, D., \& Terlevich, R. J. 1992,
\mnras, 257, 187

\reference{}
Guzm\'an, R., \& Lucey, J. 1993,
\mnras, 263, L47

\reference{}
Han, M. \& Mould, J. R. 1992, \apj, 396, 453

\reference{}
Hill, R. J. \etal\ 1997, \apjs, in press. 

\reference{}
Hernquist, L. 1990, \apj, 356, 359

\reference{}
Hjorth, J., \& Madsen, J. 1991, \mnras, 253, 703

\reference{}
Hjorth, J., \& Madsen, J. 1995, \apj, 445, 55

\reference{}
Hjorth, J., Vestergaard, M., S\o rensen, A. N., \& Grundahl, F. 1995,
\apj, 452, L17

\reference{}
Ho, L. C., Filippenko, A. V., \& Sargent, W. L. W. 1995, \apjs, 98, 477

%\reference{}
%Jacoby, G. H., Branch, D., Ciardullo, R., Davies, R. L., Harris, W. E., Pierce,
%M. J., Pritchet, C. J., Tonry, J. L., \& Welch, D. L. 1992, \pasp, 104, 599

%\reference{}
%Jacoby, G. H. 1997, in
%The Extragalactic Distance Scale,
%eds.\ M. Livio, M. Donahue, \& N. Panagia (CUP), in press

\reference{}
Jaffe, W. 1987, 
in IAU Symp.~127, Structure and Dynamics of Elliptical Galaxies, ed.\ T.
de Zeeuw (Dordrecht: Reidel), 
p.~511

\reference{}
Jerjen, H., \& Tammann, G. A. 1993, \aap, 276, 1

\reference{}
J\o rgensen, I. 1994, \pasp, 106, 967

\reference{}
J\o rgensen, I., Franx, M., \& Kj\ae rgaard, P. 1993, \apj, 411, 34

\reference{}
J\o rgensen, I., Franx, M., \& Kj\ae rgaard, P. 1995, \mnras, 276, 1341

\reference{}
J\o rgensen, I., Franx, M., \& Kj\ae rgaard, P. 1996, \mnras, 280, 167 (JFK)

%\reference{}
%Kundic, T. \etal\ 1996, \apjl, submitted

\reference{}
Landolt, A. U. 1992, \aj, 104, 340

\reference{}
Lucey, J. R., Guzm\'an, R., Carter, D., \& Terlevich, R. J. 1991,
\mnras, 253, 584

%\reference{}
%Lynden-Bell, D. \etal\ 1988, \apj, 326, 19

\reference{}
McElroy, D. B. 1995, \apjs, 100, 105

%\reference{}
%Mihalas, D., \&  Binney, J. 1981, Galactic Astronomy, 2nd ed., Freeman,
%San Francisco

\reference{}
Mould, J. \etal\ 1995, \apj, 449, 413

%\reference{}
%Mould, J. 1996, Proc.\ Heron Island Meeting, preprint

%\reference{}
%Nair, S. 1996, in Astrophysical Applications of Gravitational Lensing,
%IAU Symp.~173, eds. C.S. Kochanek \& J.N. Hewitt (Kluwer Dordrecht), p.~197

\reference{}
Pahre, M. A., Djorgovski, S. G., \& de Carvalho, R. R. 1995, \apj,
453, L17

\reference{}
Prugniel, Ph., \& Simien, F. 1994, \aap, 282, L1

\reference{}
Renzini, A., \& Ciotti, L. 1993, \apj, 416, L4

%\reference{}
%Richter, G., Longo, G., Lorenz, H., \& Zaggia, S. 1992, ESO Messenger, 68, 48

%\reference{}
%Riess, A. G., Press, W. H., \& Kirshner, R. K. 1996, astro-ph/9604143

%\reference{}
%Ruiz-Lapuente, P. 1996, \apj, 465, L83

\reference{}
Saglia, R. P., Bender, R., \& Dressler, A. 1993, \aap, 279, 75

\reference{}
Sakai, S., Madore, B. F., Freedman, W. L., Lauer, T. R., Ajhar, E. A.,
\& Baum, W. A. 1997, \apj, in press

\reference{}
Sandage, A., Saha, A., Tammann, G. A., Labhardt, L., Panagia, N.,
\& Macchetto, F. D. 1996, \apj, 460, L15

\reference{}
Sandage, A., \& Tammann, G. A. 1981, A Revised Shapley--Ames Catalog
of Bright Galaxies (Washington: Carnegie Institution) (RSA)

%\reference{}
%Schaefer, B. E. 1996, \apj, 460, L19

\reference{}
Schneider, S. 1989, \apj, 343, 94

%\reference{}
%Scorza, C., \& Bender, R. 1995, \aap, 293, 20

\reference {}
Sodemann, M., \& Thomsen, B. 1994, \aap, 292, 425

\reference{}
Sodemann, M., \& Thomsen, B. 1995, \aj, 110, 179

\reference{}
Sodemann, M., \& Thomsen, B. 1996, \aj, 111, 208

\reference{}
Statler, T. S. 1994, \aj, 108, 111

%\reference{}
%Tammann, G. A., Labhardt, L., Federspiel, M., Sandage, A.,
%Saha, A., Macchetto, F. D., \& Panagia, N. 1996, in
%Science with the Hubble Space Telescope II, eds.\
%Benvenutti, P, Macchetto, F. D., \& Schreier, E. J.,
%ST-ScI.

\reference{}
Tanvir, N. R. 1996, in
Clusters, Lensing, and the Future of the Universe, ASP Conf.\ Ser.\ 88,
eds.\ V. Trimble, \& A. Reisenegger (ASP, San Francisco), p.\ 251

\reference{}
Tanvir, N. R. 1997, in
The Extragalactic Distance Scale,
eds.\ M. Livio, M. Donahue, \& N. Panagia (CUP), in press

\reference{}
Tanvir, N. R., Shanks, T., Ferguson, H. C., \& Robinson, D. R. T.
1995, \nat, 377, 27


%\reference{}
%Thomsen, B., \& Baum, W. A. 1989, \apj, 347, 214

\reference{}
Thomsen, B., Baum, W. A., \& Hammergren, M. 1997, \apjl, submitted

\reference{}
Tonry, J. L., Blakeslee, J. P., Ajhar, E. A., \& Dressler, A. 1997, \apj, 
in press

%\reference{}
%Tremblay, B., Merritt, D., \& Williams, T. B. 1995, \apj, 443, L5

\reference{}
Tremaine, S. 1987,
in IAU Symp.~127, Structure and Dynamics of Elliptical Galaxies, ed.\ T.
de Zeeuw (Dordrecht: Reidel), 
p.~367

\reference{}
White, S. D. M. 1987,
in IAU Symp.~127, Structure and Dynamics of Elliptical Galaxies, ed.\ T.
de Zeeuw (Dordrecht: Reidel), 
p.~339

\reference{}
Worthey, G., Faber, S. M., \& Gonzalez, J. J. 1992, \apj, 398, 69

\reference{}
Yasuda, N., \& Okamura, S. 1996, \apjl, 469, L73

%\reference{}
%Yasuda, N., Fukugita, M., \& Okamura, S. 1996, \apj, submitted

\end{references}
\end{document}